# NEWS FROM LOW MASS STAR NUCLEOSYNTHESIS AND MIXING

M. Busso[a,*], E. Maiorca[a], L. Magrini[b], S. Randich[b], S. Palmerini[a], S. Cristallo[c]

*a) Department of Physics, University of Perugia and INFN, Section of Perugia*
*b) INAF, Arcetri Observatory, Florence (Italy)*
*c) Departamento de Fisica Teorica y del Cosmo, Universidad de Granada (Spain)*
*\* Corresponding author, e-mail: maurizio.busso@fisica.unipg.it*

*Abstract*

Light and intermediate nuclei as well as s-process elements have been detected in presolar grains and in evolved red giants. The abundances of some of these nuclei cannot be accounted for by canonical stellar models and require non-convective mixing below the envelope, occurring during the phases of the Red Giant Branch (RGB) and of the Asymptotic Giant Branch (AGB). Similar mechanisms appear to be necessary to account for the formation of the neutron source driving s processing. We present a short review of these phenomena and we comment on the picture that emerges from the set of available data on the evolution and nucleosynthesis in low mass stars. Our conclusions include: i) the need for deep mixing in both RGB and AGB stars; ii) the suggestion that these phenomena occur at a non-negligible velocity, possibly incompatible with diffusive processes; iii) the verification that the abundances of neutron-rich nuclei are presently increasing in the Galaxy, contrary to previous expectations and hence that the s process has new surprises to offer us; iv) the recognition of the growing importance of very low mass stars for Galactic nucleosynthesis.

## INTRODUCTION

Evolved low-mass stars show CNO isotopic ratios that are not reproduced by common evolutionary codes. It was recognized by many authors [1-5] that these anomalies derive from transport mechanisms linking the envelope to zones where partial H-burning occurs. This phenomenon has been referred to with several names; here we shall prefer the terms "extra-mixing" or "deep mixing". A few years ago a parametric study of such a process was presented [6], suitable to account for CNO abundances in presolar grains of AGB origin. The work was based on two free parameters, namely the rate of mass transport $\dot{M}$ and the temperature $T_P$ of the deepest zones reached by the circulation. It was also shown that important composition changes can occur without introducing feedbacks on the stellar luminosity, provided $T_P$ is small enough (this is expressed by saying that the difference $\Delta = \log T_H - \log T_P$ must be larger than 0.08 - 0.1, where $T_H$ is the temperature at which the maximum energy generation by H-burning is obtained).

Subsequently, physical models for extra-mixing were explored, which avoid the difficulties previously identified in mechanisms directly related to stellar rotation [7]. One of them starts from the evidence that a double diffusion can occur, as in the oceans, accompanied by transport of both temperature and chemical differences. The driving mechanism is a molecular weight inversion (see Figure 1) generated by $^3$He burning into $^4$He and 2 protons [8-11]. This process was actually discussed already many years ago [12,13] and called *thermohaline diffusion*.

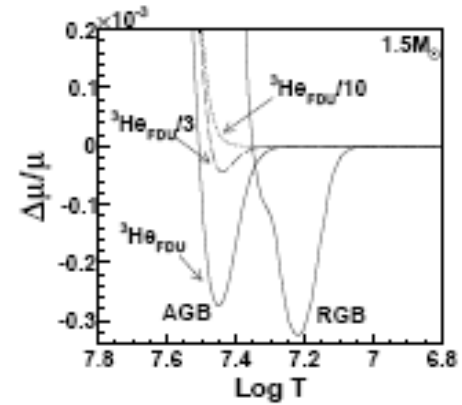

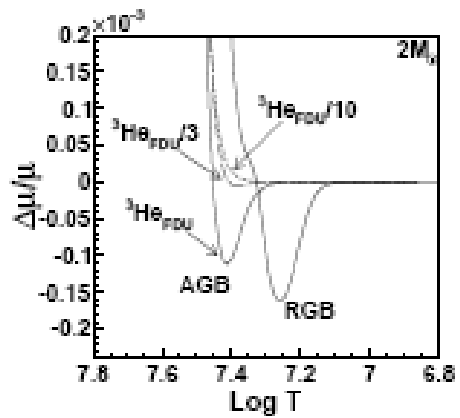

Fig. 1. The molecular weight inversion generated by the reaction $^3$He+$^3$He $\rightarrow$ $^4$He+2p in our models for a 1.5 $M_\odot$ (top) and a 2 $M_\odot$ star (bottom). The inversion exists on the RGB. Once on the AGB, its presence depends on the remaining $^3$He abundance in the envelope [22].

Another suggestion was presented by [14], using previous models by [15,16,17], according to which the magneto-hydrodynamics of the radiative regions at the base of the convective envelope in a red giant might be the site where a dynamo mechanism is established, like in the Sun. From this region magnetized bubbles lighter than the environmental material might float to the envelope promoting the chemical changes, driven by magnetic buoyancy. The magnetized domains exchange heat with the environment, thus slowing down their motion, while floating to the surface[18], but this process is inversely proportional to the area of the magnetized domains [19]. Sufficiently small bubbles or instabilities (100m in diameter or so) would travel essentially at the maximum (Alfvén) speed (Figure 2), much larger than for thermohaline mixing [14].

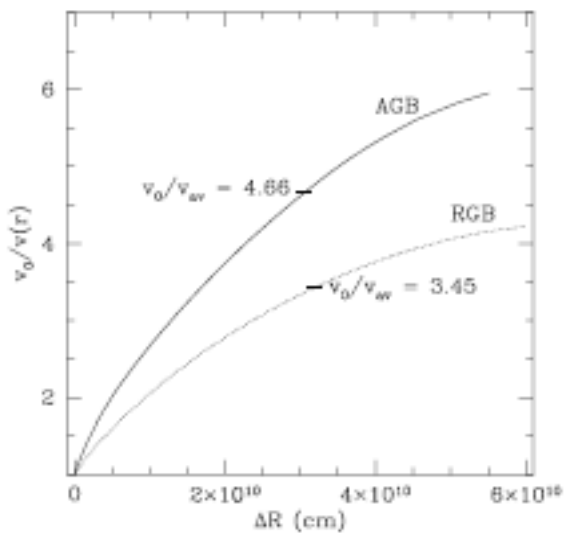

Fig. 2. The buoyancy velocity for negligible heat transfer to the environment in models of a $1.5 M_\odot$ star, during RGB and AGB stages (see [14] for details).

In [6] the role of extra-mixing on the AGB was addressed. However, calculations for the previous RGB stages were not discussed and we include them here. We also incorporate in our models several improvements in the set of reaction rates relevant for H-burning scenarios [20, 21] and we adopt recent stellar models where several physical improvements have been included [22].

A more complete discussion of our approach can be found in [23,24]. Complementary results to those shown here are included in another paper of these proceedings [25]. In that paper the attention is on intermediate-mass nuclei and on the modelling of their abundance in presolar grains. Here we want instead to consider the consequences for lithium production and destruction on the RGB and for the s-process on the AGB.

In the second section we discuss our results for the RGB stages, especially for the production and destruction of lithium. In the third section we integrate this by assuming that extra-mixing occurs also in the interpulse periods of the TP-AGB stars, below the convective envelope that penetrates the H-He discontinuity at the third dredge-up. On this basis, we discuss the new evidence emerging from observations of s-elements in young open clusters, where the abundances of the elements produced by the main s-process component considerably increase with respect to the Sun. We therefore model through extra-mixing the formation of the so-called $^{13}$C-pocket (where the main neutron source for s-processing is produced), tuning the parameters on the recent observations from open clusters. An enhanced production of n-rich nuclei, as observed in the recent Galaxy, is shown to require that the proton penetration producing the $^{13}$C-procket is more extended in very low mass stars (M = 1.2 - 1.5 $M_\odot$), remaining O-rich, than in more massive C-stars (M from 1.7 to about 3 $M_\odot$). This dependence on the initial mass is the same that is necessary to assume in order to explain CNO isotopes in red giants [25]. General conclusions are then summarized in the final section.

## PRODUCTION AND DESTRUCTION OF LITHIUM IN RGB STARS

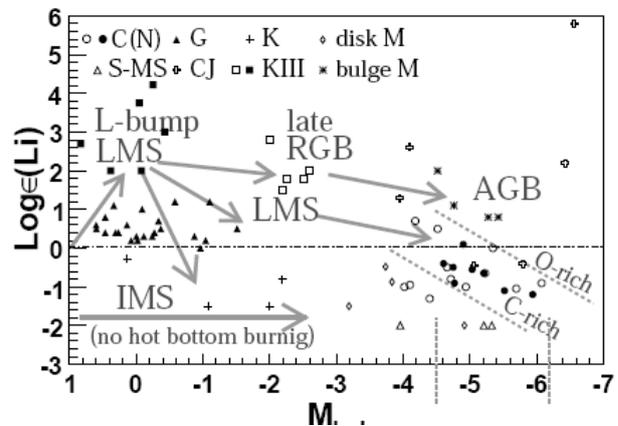

Fig. 3. Various classes of red giants with observations of Li in their photosphere. The stellar magnitudes have been determined by [26] and define a possible evolutionary history (arrows), which must include fast production on the RGB, near the bump of the Luminosity Function, and subsequent slower destruction.

Figure 3 shows a (partial) view of the abundances of lithium in red giants as a function of their luminosity, taken from [26] (see also references therein). A very small number of Li-rich stars (Log $\varepsilon$(Li) > 1.5) appears to be produced immediately after the occurrence of the first dredge-up (FDU), when a star becomes a giant. The common interpretation ascribes this to the occurrence of mixing phenomena, capable of saving to the surface the live $^7$Be produced, in the radiative layers below the envelope, by the reaction $^3$He+$^4$He. An important point however concerns the velocity of such mixing. For the typical speeds of diffusive mechanisms (cm/sec or less) the time required to carry fresh material from the H-burning zones to the envelope is of the order of at least 100 yr, much longer than $^7$Be-lifetime. Any circulation established in this way would therefore carry the Li abundance present in the envelope to hot regions, where it would be destroyed, and would restore the mixed mass not with fresh $^7$Be (that would again decay into Li), but with the products of its burning (i.e. $^4$He).

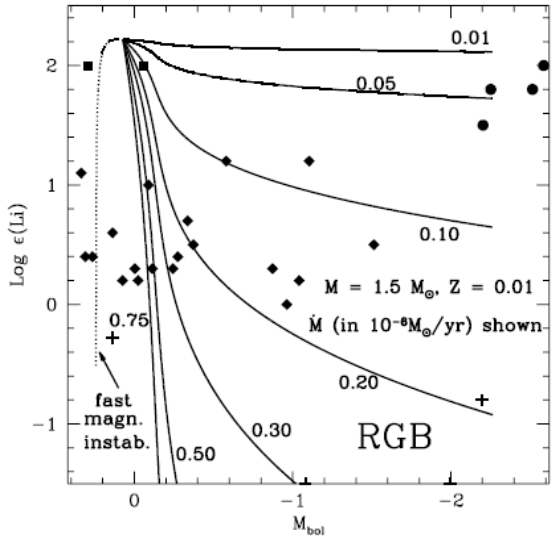

Fig. 4. Deep mixing models on the RGB yielding production or destruction of Li, depending on the rate of mass circulation. Saving $^7$Be on the surface requires faster mixing mechanisms than thermohaline diffusion.

Slow mixing is therefore bound to destroy Li. We need fast mixing to produce it, something early noticed by Cameron and Fowler [27]. As Figures 3 and 4 illustrate, the appearance is that Li is shortly produced in early RGB phases, and then largely destroyed. But this requires mixing mechanisms capable of proceeding at very different speeds. According to recent hydro-dynamical models [28, 29] this would not be the case with thermohaline diffusion, which would be too slow by orders of magnitude. As mentioned, magnetic buoyancy is instead suited, by its same nature, to produce mixing processes whose velocity depends inversely on the dimensions of the buoyant magnetized structures. Large structure will travel slowly, small ones will travel at high speed. A simple scheme in which one assumes that the instabilities gradually grow in size, slowing down in velocity, yields therefore model predictions like those of Figure 4, where the production and destruction of lithium, together with the Li observations in red giants, can all be explained rather naturally [26].

## NEWS FROM THE S-PROCESS IN THE YOUNG GALACTIC DISK

The synthesis of nuclei belonging to the main s-process component in AGB stars is known to require the activation of the $^{13}$C($\alpha$,n)$^{16}$O reaction for the production of most of the neutrons needed [30, 31]. Only a marginal second neutron flux, at higher neutron density, derives in the thermal pulses by the reaction $^{22}$Ne($\alpha$,n)$^{25}$Mg [32].

A possible scenario for producing $^{13}$C in the radiative regions of an AGB star was indicated by Cristallo et al. [22]. In this case there is not really an independent extra-mixing process introducing protons into the He-rich region. Rather, it is the convective envelope border that is assumed, more physically than in models using a bare Schwarzschild criterion, to be connected to the radiative layers in a smooth way, with convective velocities that do not drop abruptly to zero, but decrease exponentially. The region of this velocity decrease below the envelope, when the third dredge-up is formed, naturally produces a partially mixed region where a few protons remain trapped in He-rich layers. While the details of this process are different from the older parameterized models by [30, 31], the bulk results are on average the same, even if the theoretical spread in the s-process indexes found by [22] is definitely thinner.

When we were already starting to think that s-processing might not actually require a dedicated mixing mechanism to produce the $^{13}$C source, the first observations of neutron-rich elements in young open stellar clusters changed the picture once again. The general scheme of Galactic evolution for a typical neutron-rich element dominated by the s-process is shown in Figure 5, taken from [32]. There the neutrons are produced mainly by stars of masses larger than 1.5 $M_\odot$, i.e. the typical AGB stars we see today.

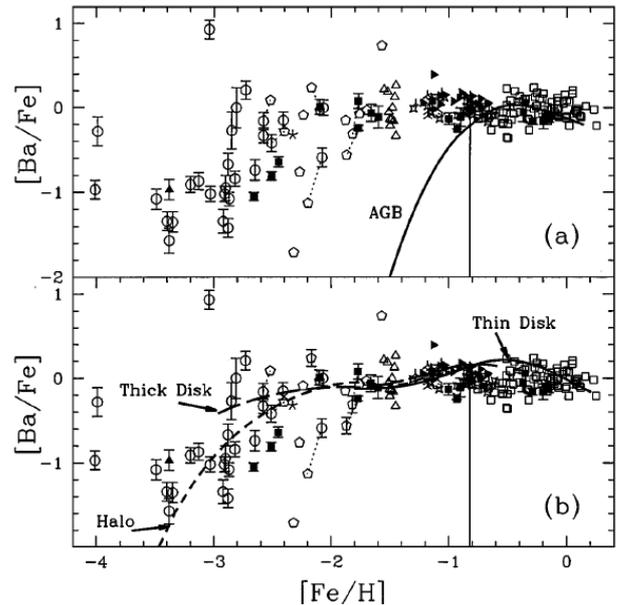

Figure 5. The evolution of Ba in the Galaxy according to [32]. Curve in the top panel: thin disk alone, AGB main component alone; curves in the bottom panel: all Galactic components, and including the r-process.

They are perfectly suitable to explain s-elements up to the epoch of the Sun's formation, and also the detail of the s-process versus r-process components in the Sun. But model curves make clear that the prediction of this scenario is that s-processing (at least for the heavy species like Ba, dominated by the main component in low mass stars) should now be dying out in the Galaxy, having already reached and passed its abundance peak.

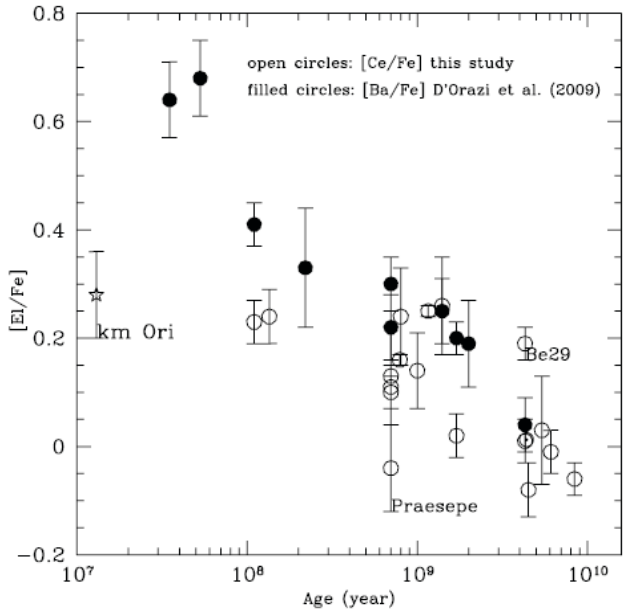

Fig. 6. The increase in [Ce/Fe] and [Ba/Fe] in open clusters younger than the Sun, according to [35].

We would therefore expect a constant or decreasing trend for the [Ba/Fe], [Ce/Fe] ratios etc. (see also [33])

However, Figure 6 demonstrates that this is not the behaviour revealed by observations in open clusters. Their stars actually show increasing abundances of neutron-rich nuclei for increasing time, or for decreasing cluster age [34,35]. A way out of this situation is to include the nucleosynthesis from very low mass stars (M < 1.5 $M_\odot$), which were barely contributing at the epoch of the solar formation.

Although detailed work on this subject is still in progress, one can show the effects of assuming that very low mass stars in their final stages have a more extended mixing below the envelope border at the third dredge-up than common (more massive) AGB stars. Let us see a rough example, using (in stars of 1.2-1.4 $M_\odot$) a $^{13}$C-pocket like the one of Figure 7, i.e. extended in mass by a factor of 3 with respect to [32]. In this way, the fit to the Ce abundances in open clusters can be improved, as displayed in Figure 8. In that Figure the dashed line presents the classical result, i.e. the sum of the r-process and of the s-process components of Ce, considering for this last part contributions only from masses larger than 1.5$M_\odot$. On the contrary, the dotted line includes, for the s-process, neutron captures induced in very low mass stars by the enlarged $^{13}$C pocket of Figure 7. Even in this over-simplified example it is clear that the new hypotheses provide exactly the contributions that were missing.

Similar trends and a similar interpretation also apply to lighter s-process elements, like Y, where however the main s-process component does not suffice and a secondary production by intermediate masses and/or by massive stars (weak component) is needed (see e.g. [36]).

An important check of the suggestions here advanced on the extension and effects of extra-mixing processes should be provided by fluorine data. Indeed, $^{19}$Fe is produced in AGB stars mainly from the same $^{13}$C pocket where s-process nuclei are synthesized. Then it is dredged-up in the H-rich layers and in the envelope. Its abundance can be further modified in the H-burning shell and in any deep circulation affecting proton captures, so that its final abundance might offer a way to check together deep-mixing models for He- and H-burning zones. Due to these interesting properties, we have now a program for observing fluorine in the same open clusters for which the s-element abundances were derived.

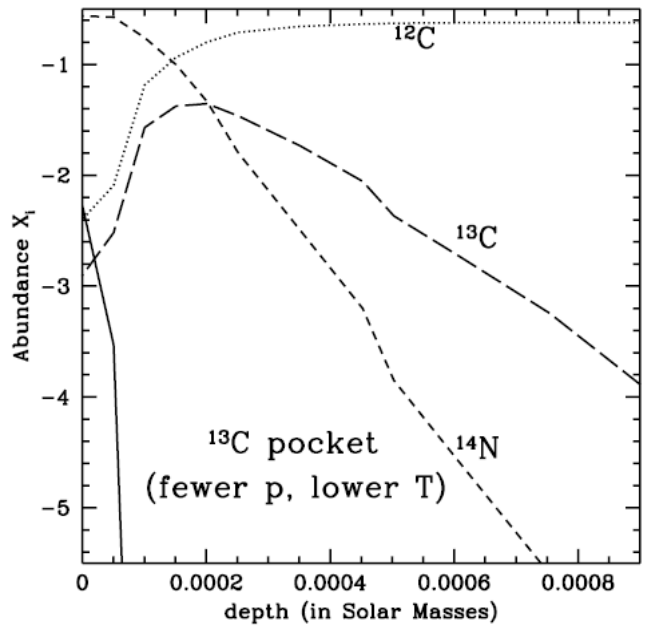

Figure 7. A $^{13}$C-procket obtained by extending the profile of the penetrated protons below the envelope by a factor of 3 as compared to [32].

## PRELIMINARY CONCLUSIONS

The nucleosynthesis pattern of low mass stars, both in their RGB and in their AGB phases requires the activation of non-convective mechanisms of mixing for transporting materials across the inner border of the convective envelope.

On the RGB, the observations require only a shallow type of mixing, for which the double diffusive process called *thermohaline mixing* seemed until recently to be adequate; doubts on its efficiency advanced in the past year now require independent verifications of the models by other groups.

Thermohaline diffusion is however certainly incapable of mixing fast enough to account for the minority of stars that show important episodes of Li production (Log ε(Li) > 1.5). For them transport phenomena based on faster mechanisms are needed and we showed that one such mechanism might be the rapid buoyancy of small (< 1Km in size) magnetic instabilities from toroidal magnetic flux tubes.

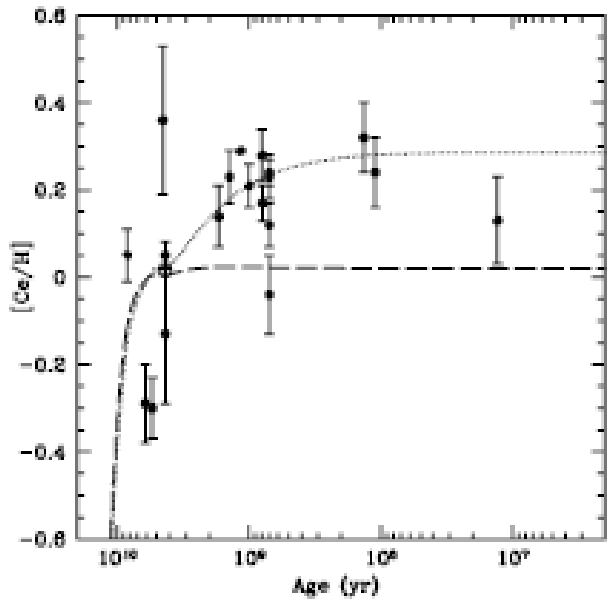

Figure 8. Dotted line: the Galactic chemical evolution of Ce, including the contribution of low masses (M < 1.5 M$_\odot$) and assuming for them a $^{13}$C pocket as in Figure 7. Dashed line: the same, with the old scenario for the $^{13}$C pocket. Both curves have the r-process contribution added.

Thermohaline mixing is also incapable to account for the deep circulation required by $^{26}$Al observations in presolar grains of AGB origin, as well as for the formation of the $^{13}$C neutron source below the convective envelope at the occurrence of TDU.

Recent observations of neutron-capture elements in stars of open clusters reveal that the s-process nucleosynthesis ensuing from the formation of the above neutron source calls for contributions of very low mass stars, born shortly after the epoch of the Galactic disc formation and allowing for a deeper penetration of protons below the envelope than required for more massive (M > 1.5 M$_\odot$) stars.

This last indication on effective extra-mixing processes in very low mass stars from neutron-capture nuclei is qualitatively similar to the requirements imposed by the observations of lighter nuclei in red giants.

One should now verify whether some indications on an enhanced production of s-process nuclei by low mass stars can be identified also in the Sun. Although at the epoch of the Sun's formation these contributions should be in general minimal (few such stars would have contributed, for a question of time) and the bulk solar distribution of neutron capture elements should be unchanged, some effects might be visible in the more subtle isotopic anomalies imprinted by a possible last-minute contamination of the solar nebula, as suggested by the abundances of some short-lived radioactive nuclei of neutron-capture origin, like $^{60}$Fe and $^{205}$Pb [37].

In general the above findings mutually agree in depicting a scenario in which all the nucleosynthesis signatures of low mass stars are dominated by the action of elusive, non convective transport phenomena, linking the convective envelope with more internal regions affected by thermonuclear reactions.